\documentclass[conference]{IEEEtran}
\IEEEoverridecommandlockouts
\usepackage{cite}
\usepackage{amsmath,amssymb,amsfonts}
\usepackage{graphicx}
\usepackage{algorithm, tabularx}
\usepackage{textcomp}
\usepackage{xcolor}
\usepackage{epsfig}
\usepackage{epstopdf}
\usepackage{balance}
\usepackage{algpseudocode}
\usepackage[printonlyused]{acronym}
\usepackage{physics}
\acrodef{bs}[BS]{base station}
\acrodef{mimo}[MIMO]{multiple-input multiple-output}
\acrodef{thz}[THz]{Terahertz}
\acrodef{omp}[OMP]{orthogonal matching pursuit}
\acrodef{amp}[AMP]{approximate message passing}
\acrodef{snr}[SNR]{signal-to-noise ratio}
\acrodef{los}[LoS]{line-of-sight}
\acrodef{nlos}[NLoS]{non-line-of-sight}
\acrodef{aoa}[AoA]{angle of arrival}
\acrodef{ula}[ULA]{uniform linear array}
\acrodef{nmse}[NMSE]{normalized mean-square error}
\acrodef{cnn}[CNN]{convolutional neural network}
\acrodef{ad}[AD]{angular domain}
\acrodef{pd}[PD]{polar domain}
\acrodef{mmimo}[mMIMO]{massive multiple-input multiple-output}
\newcommand{\FGR}[1]{Fig.~\ref{#1}}

\def\BibTeX{{\rm B\kern-.05em{\sc i\kern-.025em b}\kern-.08em
    T\kern-.1667em\lower.7ex\hbox{E}\kern-.125emX}}
\begin{document}

\makeatletter
\newcommand{\multiline}[1]{%
  \begin{tabularx}{\dimexpr\linewidth-\ALG@thistlm}[t]{@{}X@{}}
    #1
  \end{tabularx}
}
\makeatother

\title{Channel Estimation Using RIDNet Assisted OMP for Hybrid-field THz Massive MIMO Systems}

\author{\IEEEauthorblockN{Hasan Nayir\IEEEauthorrefmark{1}\IEEEauthorrefmark{2}\IEEEauthorrefmark{3}, Erhan Karakoca\IEEEauthorrefmark{1}\IEEEauthorrefmark{2}\IEEEauthorrefmark{3}, Ali Görçin\IEEEauthorrefmark{3}\IEEEauthorrefmark{4}, Khalid Qaraqe\IEEEauthorrefmark{1}}

\IEEEauthorblockA{\IEEEauthorrefmark{1} Department of Electrical and Computer Engineering, Texas A\&M University at Qatar, Doha, Qatar}

\IEEEauthorblockA{\IEEEauthorrefmark{2} Department of Electronics and Communication Engineering, Istanbul Technical University, {\.{I}}stanbul, Turkey}

\IEEEauthorblockA{\IEEEauthorrefmark{3} Communications and Signal Processing Research (HİSAR) Lab., T{\"{U}}B{\.{I}}TAK B{\.{I}}LGEM, Kocaeli, Turkey}

\IEEEauthorblockA{\IEEEauthorrefmark{4} Department of Electronics and Communication Engineering, Yildiz Technical University, {\.{I}}stanbul, Turkey\\
Emails: \texttt{\{nayir20, karakoca19\}@itu.edu.tr, agorcin@yildiz.edu.tr,}\\
\texttt{{khalid.qaraqe@qatar.tamu.edu}}}} 
\maketitle

\begin{abstract}
The terahertz (THz) band radio access with larger available bandwidth is anticipated to provide higher capacities for next-generation wireless communication systems. However, higher path loss at THz frequencies significantly limits the wireless communication range. Massive multiple-input multiple-output (mMIMO) is an attractive technology to increase the Rayleigh distance by generating higher gain beams using low wavelength and highly directive antenna array aperture. In addition, both far-field and near-field components of the antenna system should be considered for modelling THz electromagnetic propagation, where the channel estimation for this environment becomes a challenging task.
This paper proposes a novel channel estimation method using a recursive information distillation network (RIDNet) together with orthogonal matching pursuit (OMP) for hybrid-field THz mMIMO channels, including both far-field and near-field components. The simulation experiments are performed using the ray-tracing tool. The results indicate that the proposed RIDNet-based method consistently provides lower channel estimation errors compared to the conventional OMP algorithm for all signal-to-noise ratio (SNR) regimes, and the performance gap becomes higher at low SNR regimes. Furthermore, the results imply that the same error performance of the OMP can be achieved by the RIDNet-based method using a lower number of RF chains and pilot symbols.

\end{abstract}


\section{Introduction}

The \ac{thz} band provides ultra-high bandwidth to satisfy the increasing data rate requirements of next-generation wireless communication systems such as 6G \cite{sarieddeen2021overview}. However, the \ac{thz} frequency band suffers from high molecular absorption and spreading losses, which limits wireless propagation resulting in lower communication coverage. The advent of \ac{mmimo} structures has enabled the generation of high-gain beams, which can potentially overcome the higher path loss issue in the \ac{thz} band \cite{akyildiz2018combating}. Accurate channel estimation is required to generate high-gain beams; however, a reduced RF chain along with the limited number of pilots makes this a challenging task. 





Electromagnetic propagation characteristics can be divided into two categories, namely near-field and far-field \cite{lu2022double}. The Rayleigh distance, which can be calculated by dividing the square of the antenna array aperture by the wavelength, determines the boundary between the far-field and near-field. If the distance between the base station and the signal source is greater than the Rayleigh distance, it indicates that the signal source comes from the far-field region and therefore propagates as a plane wave. Otherwise, the signal source is in the near-field region, and the electromagnetic wave propagates as a spherical wave.


The number of antennas in traditional cellular communication systems is relatively small, and the Rayleigh distance is only a few meters, so near-field components can be neglected. However, the Rayleigh distance is expected to increase significantly in 6G systems due to a large number of antennas. For example,  researchers have designed a $2m\times 3m$ array with $3200$ antenna elements operating at $2.4$GHz in \cite{246282}. The Rayleigh distance of this antenna array is around $200m$, which is greater than a typical 5G cell. In addition, the Rayleigh distance of the antenna array operating at $300$GHz and having an antenna aperture of $0.1m$ is approximately $20m$, which provides relatively higher coverage in \ac{thz} systems. Therefore, hybrid-field channel models incorporating far-field components along with near-field components should be considered to characterize the wireless communication channel \cite{wei2021channel}. 

Matrix transformation has been utilized in the literature for compressive sensing-based channel estimation with lower pilot overhead \cite{cui2022channel}. However, traditional compressive sensing-based methods have limited channel estimation performance, especially when the number of RF chains is low compared to the number of antennas and the number of pilot symbols is limited. Deep learning methods have been used recently to support traditional methods and improve the accuracy of channel estimation \cite{wei2019amp, hu2022prince, he2018deep, ma2020sparse, chen2021hybrid, jin2019channel}. For example, a two-stage channel estimation method is presented in \cite{wei2019amp, hu2022prince}, where \ac{amp} is used as a coarse estimation in the first stage, followed by \ac{cnn} structure for channel feature extraction to decrease the channel estimation error. Inspired by these works, in this paper, we propose a hybrid-field channel estimation mechanism using a recursive information distillation network (RIDNet) to improve the estimation performance of the \ac{omp} algorithm, especially at low \ac{snr}s. The proposed method successfully compensates for errors caused by not only having the number of RF chains lower than the number of antennas but also the number of limited pilot symbols due to \ac{mmimo} systems.

The performance of the proposed channel estimation scheme is evaluated using the ray tracing tool in Matlab, where the Sketchup program is utilized to model a hybrid-field environment including both near-field and far-field components. The RIDNet-assisted OMP scheme consistently provides lower channel estimation error for all SNR regimes, and the performance improvement increases as the SNR decreases. The proposed scheme achieves lower error performance using 8 RF chains compared to the OMP scheme using 12 and 16 RF chains. Similarly, the same error performance can be achieved by the proposed scheme using a lower number of pilot symbols.


Throught this paper, $\mathbf{x}$ and $\mathbf{X}$ denote a vector and matrix, respectively. $\lVert x \rVert$ represents the norm of $\mathbf{x}$. $\mathbb{C} \mathcal{N}(\mu,\,\sigma_{n}^{2})$ identify the probability density function of the complex Gaussian distribution with mean $\mu$ and $\sigma_{n}^{2}$ variance. 

The rest of this paper is organized as follows: Section II introduces hybrid-field \ac{thz} \ac{mmimo} channel that includes path loss and sparse representation of the channel. In Section III, the proposed \ac{omp}-RIDNet-based hybrid-field channel estimation method is detailed. Finally, simulation results and conclusions are elaborated in Section IV and Section V, respectively. 

\section{System Model}
In this section, we present the proposed uplink channel estimation system model for THz \ac{mmimo} OFDM communication systems with multiple time slots allocated to each user, as depicted in \FGR{fig:main_figure}. The \ac{bs} is equipped with a \ac{ula} of $N_{RF}$ RF chains and $N$ antenna elements such that there is a spacing of $d=\frac{\lambda}{2}$ between two consecutive antenna elements, where $\lambda$ represents the carrier wavelength. In addition, a hybrid precoding architecture is utilized in the \ac{bs} to improve energy efficiency by reducing the number of RF chains. We assume that $M$ subcarriers are served simultaneously to $K$ users, and orthogonal pilot symbols are transmitted to the \ac{bs} for channel estimation.

\begin{figure*}
    \centering
    \includegraphics[width=0.95\linewidth]{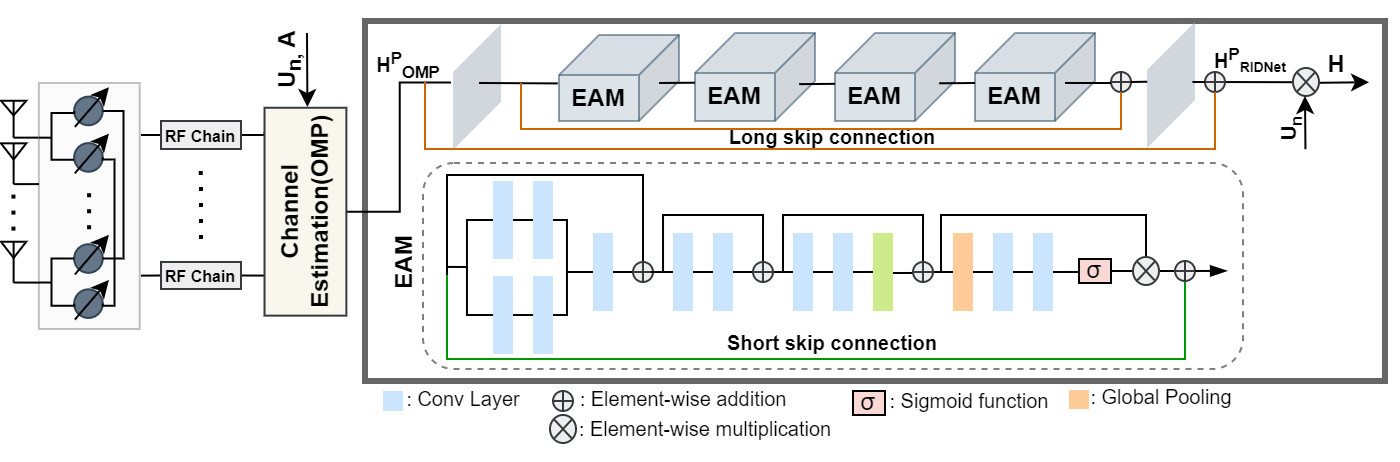}
    \caption{The proposed channel estimation scheme based on OMP-RIDNet structure.}
    \label{fig:main_figure}
\end{figure*}


Let $y_{m,p} \in \mathbb{C}^{N_{RF}\times1}$ represent the uplink received signal
\begin{equation}
\mathbf{y}_{m, p}=\mathbf{W}_{p} h_{m} s_{m, p}+\mathbf{W}_{p} \mathbf{n}_{m, p},
\end{equation}
where $s_{m, p}$ is the transmit pilot symbol at the $m$-th subcarrier in time slot $p$, $n_{m,p} \in \mathbb{C}^{N \times 1}$ denotes a complex Gaussian noise with the distribution function of $\mathcal{C} \mathcal{N}\left(0, \sigma^{2} \mathbf{I}_{N}\right)$ and $\mathbf{W}_{p} \in \mathbb{C}^{N_{RF}\times N}$ represents an analog combining matrix containing the phase coefficients of the phase shifters. Also, $h_{m}$ indicates the hybrid-field channel between the \ac{bs} and the user at the $m$-th subcarrier. The overall received signal from time slot $1$ to $P$ at the \ac{bs} for the $m$-th subcarrier without the pilot symbols can be expressed as
\begin{equation}
\mathbf{y}_{m}=\left[\mathbf{y}_{m, 1}^{T}, \cdots, \mathbf{y}_{m, P}^{T}\right]^{T}=\mathbf{W} \mathbf{h}_{m}+\mathbf{n}_{m},
\end{equation}
where $\mathbf{n}_{m}=\left[\mathbf{n}_{m, 1}^{T} \mathbf{W}_{1}^{T}, \cdots, \mathbf{n}_{m, P}^{T} \mathbf{W}_{P}^{T}\right]^{T} \in \mathbb{C}^{QN_{RF}\times Q}$, and $\mathbf{W}=\left[\mathbf{W}_{1}^{T}, \cdots, \mathbf{W}_{P}^{T}\right]^{T} \in \mathbb{C}^{QN_{RF}\times N}$ identify the overall noise and analog combining matrix for $P$ time slots, respectively. Assume that $\mathbf{W}$ is generated randomly using the uniform random distribution $\mathcal{U}(-1,1)$ and normalized with $1/\sqrt{N}$.



\subsection{Hybrid-Field THz Massive MIMO Channel Model}


Far-, near-, and hybrid-field channel models are described in this section. A Rayleigh distance is determined by $D_{R}=\frac{2D^2}{\lambda}$, where $D$ and $\lambda$ represent the antenna array aperture and the wavelength of the transmitted signal, respectively. Assuming that \ac{bs} is equipped with \ac{ula} with $N$ antenna elements, antenna array aperture $D$ is equal to $N\frac{\lambda}{2}$.

The Rayleigh distance determines whether a channel includes far- or near-field propagation components. When a signal travels more than the Rayleigh distance, the signal propagates as a plane wave characterized by the far-field propagation model. Otherwise, the near-field propagation model characterizes the radiation pattern, where the signal propagates as a spherical wave. Since the Rayleigh distances are small in 5G and previous communication systems, the far-field model having only plane wave propagation is sufficient to obtain the radiation field. However, the Rayleigh distance of \ac{thz} \ac{mmimo} communication systems necessitates that both far- and near-field approaches should be considered using a hybrid approach. While there is a direct link between the \ac{bs} and a user in the far-field region, there may also be scattered signals from the same user in the near-field region. Therefore, a hybrid-field channel vector at the $m$-th subcarrier containing both far-field and near-field components can be expressed as \cite{wei2021channel}


\begin{equation}\label{hybrid_field_channel_model}
\begin{aligned}
\mathbf{h}_{m}=\sqrt{\frac{N}{L}}\left(\sum_{l_{\mathrm{f}}=1}^{L_{f}} \alpha_{l_{\mathrm{f}}} \mathbf{a}\left(\theta_{l_{\mathrm{f}}}\right)
+\sum_{l_{\mathrm{n}}=1}^{L_{n}} \alpha_{l_{\mathrm{n}}} \mathbf{b}\left(\theta_{l_{\mathrm{n}}}, r_{l_{\mathrm{n}}}\right)\right)
\end{aligned},
\end{equation}
where $L$, $L_{f}$, and $L_{n}$ represent the number of all path, far-field, and near-field components, respectively. $\alpha_{l_{f}}$ and $\theta_{l_{f}}$ denote the channel complex gain and the angle of the $l_{f}$-th path for the far-field. $\alpha_{l_{n}}$, $\theta_{l_{n}}$, and $r_{l_{n}}$ denote the channel complex gain, the angle, and the distance of the $l_{n}$-th path for the near-field, respectively. The array steering vector in the far-field region can be expressed as
\begin{equation}
\mathbf{a}\left(\theta_{l_{f}}\right)=\frac{1}{\sqrt{N}}\left[1, e^{-j \frac{2 \pi d}{\lambda} \cos \theta_{l_{f}}}, \ldots, e^{-j(N-1) \frac{2 \pi d}{\lambda} \cos \theta_{l_{f}}}\right]^{H},
\end{equation}
where $d$ is the distance between antenna elements in the \ac{ula} and assumed to be $\lambda/2$.
Furthermore, the array steering vector in the near-field region varies with respect to the distance between the \ac{bs} and the user or the scatter in addition to the angle of arrival as follows

\begin{equation}
\mathbf{b}\left(\theta_{l_{n}}, r_{l_{n}}\right)=\frac{1}{\sqrt{N}}\left[e^{-j \frac{2 \pi}{\lambda}\left(r_{l_{n}}^{(1)}-r_{l_{n}}\right)}, \ldots, e^{-j \frac{2 \pi}{\lambda}\left(r_{l_{n}}^{(N)}-r_{l_{n}}\right)}\right]^{H},
\end{equation}
where $r_{l_{n}}$ describes the distance between $l_{n}$-th scatter and the center of the antenna array and changes geometrically according to each element $z$ in the antenna array as follows: 
\begin{equation}
r_{l_{n}}^{(z)}=\sqrt{r_{l_{n}}^{2}+\Delta_{z}^{2} d^{2}-2 r_{l_{n}} \Delta_{z} \theta_{l_{n}} d },
\end{equation}
where $\Delta_{z}=\frac{2z-N-1}{2}$ with $z=1,2,...,N$.

\subsubsection{Path Loss}
Aside from spreading loss, molecular absorption is added to the total path loss at \ac{thz} frequencies due to the small wavelength.
In communication systems, there may be \ac{los} and \ac{nlos} links between the transmitter and receiver, where the channel gain expressions are different.
In the case of an \ac{los} uplink transmission, the channel gain can be calculated as \cite{chaccour2022can}

\begin{equation}\label{los_complex_gain}
\alpha^l = \frac{c}{4 \pi f_{m} r_{l}} e^{-\frac{k(f_{m}) r_{l}}{2}} e^{-j 2 \pi f_{m} \tau_{l}^{L}},
\end{equation}
where $c$, $k(f_{m})$, $f_{m}$, and $r_{l}$ denote the speed of the light, molecular absorption coefficients of the channel at \ac{thz} band, the $m$-th subcarrier frequency, and the distance between the user and the BS, respectively. However, a \ac{los} link may not always be available, and the information may be carried through the signals reflected or scattered from the environment. In the case of an \ac{nlos} uplink transmission, the channel gain can be expressed as \cite{moldovan2014and}
\begin{multline}\label{nlos_complex_gain}
   \alpha_{l}^{N}=\frac{c}{4 \pi f_{m}\left(r_{l}^{(1)}+r_{l}^{(2)}\right)} e^{\left(-\frac{k(f_{m})\left(r_{l}^{(1)}+r_{l}^{(2)}\right)}{2}\right)}\\
   \times R(f_{m}) e^{-j 2 \pi f_{m} \tau_{l}^{N}}, 
\end{multline}
where $r_{l}^{(1)}$ represents the distance between the user and a scatter location while $r_{l}^{(2)}$ denotes the distance between a scatter and the BS. $R(f_{m})=\gamma_{n, u}(f) \rho_{n, u}(f)$ denotes the reflection coefficient, where $\gamma_{n, u}(f) \approx -\exp \left(\frac{-2 \cos \left(\psi_{n, u}\right)}{\sqrt{\eta(f)^{2}-1}}\right)$ is the Fresnel reflection coefficient, where $\eta(f)$ and $\psi_{n, u}$ represent the refractive index and the angle of the incident signal to the reflector, respectively. $\rho_{n, u}(f)=\exp \left(-\frac{8 \pi^{2} f^{2} \sigma^{2} \cos ^{2}\left(\psi_{n, u}\right)}{c^{2}}\right)$ is the Rayleigh factor that characterizes the roughness effect, where $\sigma$ is the surface height standard deviation.

\subsubsection{Sparse Representation of THz massive MIMO Channel}
There have been a number of matrix transformation approaches proposed to reduce pilot overhead for \ac{mmimo} systems in addition to sparse channel representations providing a low overhead solution for channel estimation. This section presents the \ac{pd} matrix transformation approach used to represent the hybrid-field channel sparsely, where both far-field and near-field propagation components are simultaneously modeled.

A DFT matrix can be used for a channel model with only far-field components; however, it is not suitable for a channel model with hybrid-field components since the DFT matrix has only angle-related information and does not include the distance-dependent information of the near-field channel. As a result, the near-field channel is not sparse in the \ac{ad}. In this study, polar-domain transformation is used to represent the hybrid-field channel as sparse. The polar-domain representation of the channel can be expressed as
\begin{equation}
\mathbf{h}_{m}=\mathbf{U_{n} h}_{m}^{\mathcal{P}},
\end{equation}
where $\mathbf{U_{n}} \in \mathbb{C}^{N \times S}$ and $S$ are the number of sampled far and near-field steering vectors in the \ac{pd}, respectively. The \ac{pd} matrix $\mathbf{U_{n}}$ can be represented as
\begin{equation}
\begin{aligned}
\mathbf{U_{n}}=\left[\mathbf{b}\left(\theta_{1}, r_{1}^{1}\right), \ldots, \mathbf{b}\left(\theta_{1},\right.\right.&\left.r_{1}^{S_{1}}\right), \ldots, \\
&\left.\mathbf{b}\left(\theta_{N}, r_{N}^{1}\right), \ldots, \mathbf{b}\left(\theta_{N}, r_{N}^{S_{N}}\right)\right],
\end{aligned}
\end{equation}
where $s_{n}=1,2,\ldots,S_{n}$ denotes the samples of distance at each angle. Note that the columns of $\mathbf{U_{n}}$ contain samples at each angle and distance. The \ac{pd} transform accounts for both angle and distance information of all components, so the energy spread the effect of near-field components in the \ac{ad} is also eliminated.


\section{OMP and RIDNet for Hybrid-Field THz Massive MIMO Channel Estimation}
In this section, the proposed \ac{omp}-RIDNet method is described for the hybrid-field channel estimation. The method consists of two stages (i.e., \ac{omp} and RIDNET). \ac{omp} algorithm strives for better performance, especially at low SNR regions. Thus, the RIDNet further refines the estimate after a coarse estimation with \ac{omp} and provides the desired result.
\subsection{OMP Stage}
Channel estimation can be performed with the low pilot overhead using compressive sensing algorithms such as \ac{omp} in sparse channels. The pseudo-code of the \ac{omp} is described in Algorithm \ref{alg:omp_algorithm}, where $\mathbf{Y}, \mathbf{W}, \mathbf{U_{n}}, N$, and $M$ are defined above, and $T$ represents the number of iterations. During the estimation of the hybrid-field components, the matrix $\mathbf{A} = \mathbf{W}\mathbf{U_{n}}$ is used as a sensing matrix. The correlation between $\mathbf{A}$ and the residual matrix $\mathbf{R}$ is calculated at each iteration. Then, the highest correlative index is stored as an element of vector $s$, which is updated after each iteration. Using the least square algorithm, we then obtain the estimated hybrid-field sparse channel matrix in the \ac{pd}.

\begin{algorithm}
\caption{Estimation of the sparse hybrid-field channel}\label{alg:omp_algorithm}
\begin{algorithmic}[1]
\Function{OMP}{$\mathbf{Y}, \mathbf{W}, \mathbf{U_{n}}, T, N, M$}
\State $\mathbf{A} = \mathbf{W}\mathbf{U_{n}}$
\State $\mathbf{R}=\mathbf{Y}$
\For{$i\gets 1: T$}
    \State $k^{*}=\operatorname{argmax}\left\|\mathbf{A}^{H}(:, k) \mathbf{R}\right\|_{2}^{2}$
    \State $s=s\cup k^{*}$
    \State $\hat{\mathbf{H}}^{\mathcal{P}}=\mathbf{0}_{N \times M}$
    \State $\hat{\mathbf{H}}^{\mathcal{P}}\left(s\right)=\mathbf{A}^{H}\left(:, s\right) \mathbf{Y}$
    \State $\mathbf{R}=\mathbf{Y}-\mathbf{A} \hat{\mathbf{H}}^{\mathcal{P}}$
\EndFor
\EndFunction
\end{algorithmic}
\end{algorithm}
As the output of Algorithm \ref{alg:omp_algorithm}, we obtain the estimated sparse hybrid-field channel.


\subsection{RIDNet Stage}\label{ridnet_stage}
In the second stage, a feature attention-based \ac{thz} hybrid-field channel denoising network called RIDNet is utilized  further to improve the channel estimation of the \ac{omp} algorithm. The second stage consists of three main steps: feature extraction, feature learning residual on the residual module, and reconstruction.

First of all, the coarse estimated $\hat{\mathbf{H}}^{\mathcal{P}}$ using the OMP algorithm passes through the feature extraction module that includes only one convolutional layer. Thus, the initial features $f_{0}$ are obtained from noisy input $\hat{\mathbf{H}}^{\mathcal{P}}$
\begin{equation}
f_{\hat{\mathbf{H}}^{\mathcal{P}}_{0}}=f_{ConvE}(\hat{\mathbf{H}}^{\mathcal{P}}),
\end{equation}
where $f_{Conv}(\cdot)$ is the convolution operator. Next, $f_{\hat{\mathbf{H}}^{\mathcal{P}}_{0}}$ goes through the feature learning residual on the residual module, which is created by cascading the EAM modules.
\begin{equation}
f_{\hat{\mathbf{H}}^{\mathcal{P}}_{r}}=M_{fl}\left(\hat{\mathbf{H}}^{\mathcal{P}}_{0}\right),
\end{equation}
where $f_{\hat{\mathbf{H}}^{\mathcal{P}}_{r}}$ are the learned features and $M_{fl}(\cdot)$ is the main feature learning on the residual component. Then, the output features $f_{\hat{\mathbf{H}}^{\mathcal{P}}_{r}}$ from the final layer go through the reconstruction module $f_{ConvR}(\cdot)$, which consists of a single convolutional layer. 
\begin{equation}
\hat{y}=f_{ConvR}\left(f_{\hat{\mathbf{H}}^{\mathcal{P}}_{r}}\right)
\end{equation}
As depicted in \FGR{fig:main_figure}, input data is summed with $\hat{y}$ via a long skip connection, and denoised estimated channel $\hat{\mathbf{H}}_{R}^{\mathcal{P}}$ is obtained as output. For $N$ training pairs in each batch, $\{\hat{\mathbf{H}}^{\mathcal{P}}, {\mathbf{H}}^{\mathcal{P}}\}_{i=1}^{N}$, where $\hat{\mathbf{H}}^{\mathcal{P}}$ is the noisy input channel that is output of the OMP algorithm and ${\mathbf{H}}^{\mathcal{P}}$ is the ground truth, the loss function can be expressed as
\begin{equation}
L(\mathcal{W})=\frac{1}{N} \sum_{i=1}^{N}\left\|\operatorname{RIDNet}\left(\hat{\mathbf{H}}^{\mathcal{P}}_{i}\right)-{\mathbf{H}}^{\mathcal{P}}_{i}\right\|_{1},
\end{equation}
where $\mathcal{W}$ denotes the network parameters learned.


There are four EAMs in the RIDNet model. First, the input features are divided into two branches, and each passes through two convolution layers, then concaneted and passed through one more convolution layer. Furthermore, after learning the features using two convolution layers, compression is performed with a total of three convolution layers, two of which are $3\times3$ kernel size, and the third is $1\times1$ kernel size. Also, for channel estimation denoising problems, channel features are generally treated equally, but this may not be appropriate. When we feed the compressed data directly to the convolutional layer, only local information will be used. Therefore, a global average pooling is applied before the convolutional layer to get the statistics of the data. Moreover, after global average pooling, there are two convolutional layers and the sigmoid is used as an activation function in the second convolutional layer. Finally, the output of the convolutional layer containing the sigmoid is multiplied by the input of the global average pooling layer and summed with the input of the EAM module through the short skip connection.

\section{Simulation Results}
This section presents the performance evaluation of the proposed OMP-RIDNet method through simulation experiments for \ac{thz} \ac{mmimo} systems.

\subsection{Simulation Scenario}

It is hard to obtain real measurement data for massive antenna systems in the \ac{thz} band due to the hardware limitations. However, thanks to simulators such as NYUSIM \cite{ju2019millimeter}, TeraMIMO \cite{tarboush2021teramimo}, Wireless InSite \cite{hur2016proposal}, and Matlab, \ac{thz} \ac{mmimo} channels can be accurately modeled using the ray tracing technique. 

\begin{figure}[t!]
    \centering
    \includegraphics[width=0.95\linewidth]{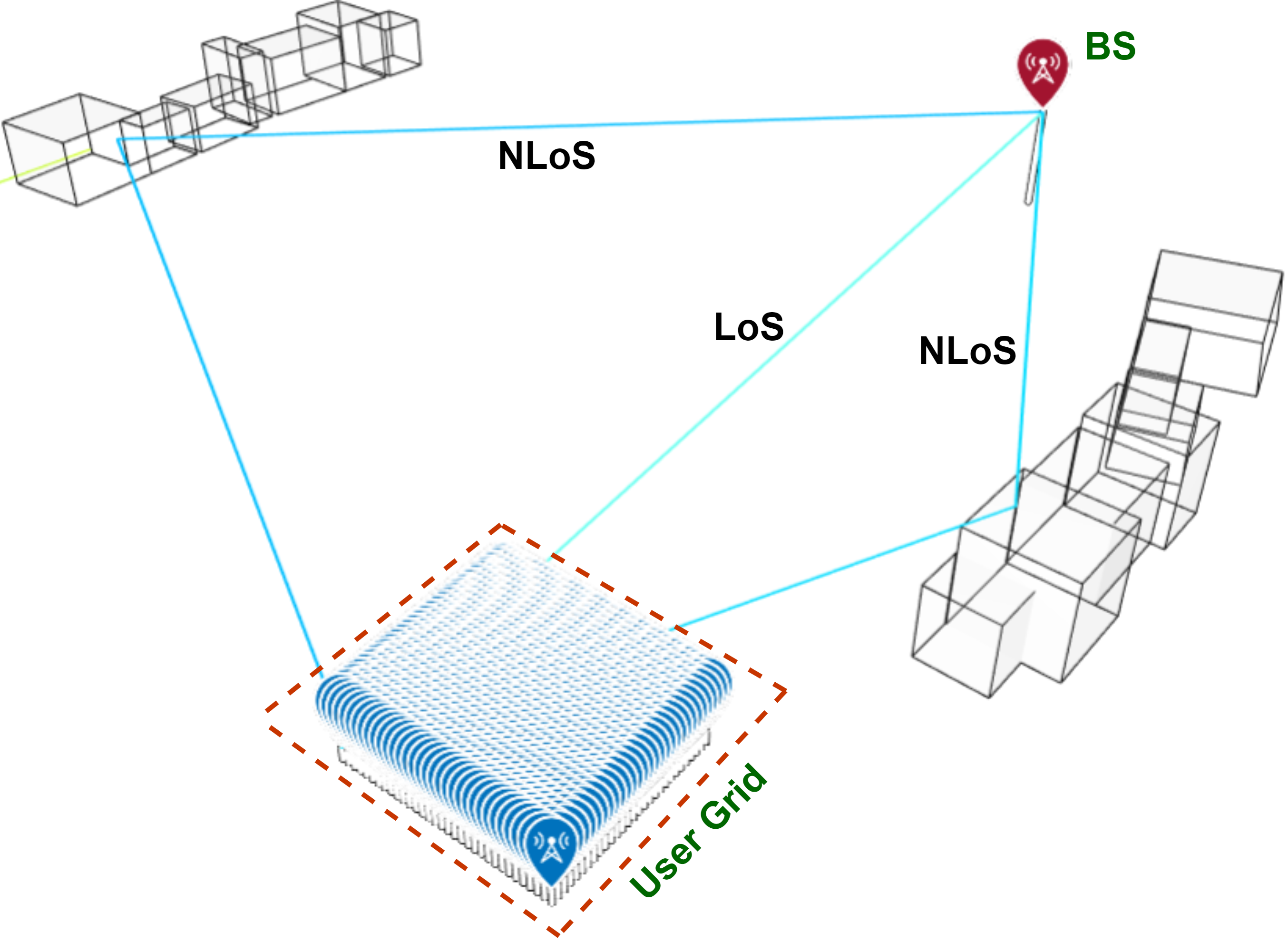}
    \caption{The simulation environment}
    \label{simulation_environment}
\end{figure}

In this study, an outdoor environment consisting of one \ac{bs}, selected 1000 measurement points on a user grid, and different shapes and sizes objects representing buildings are modeled using Sketchup as shown in \FGR{simulation_environment}. This Sketchup  environment is imported into Matlab to be utilized by the ray tracing tool for obtaining the information of \ac{los} and \ac{nlos} paths such as \ac{aoa} path delays, the phase, and the propagation distance. The \ac{los} and \ac{nlos} path properties are used in (\ref{hybrid_field_channel_model}), (\ref{los_complex_gain}), and (\ref{nlos_complex_gain}) to construct the hybrid-field channel matrix. In particular, a \ac{ula} of $N=256$ elements is used as the antenna in the BS. In a typical communication environment, the number of multipath components depends on the user locations. 
For a stationary Tx on a particular location, measurements are taken from 1000 different receiver locations to make an analysis closer to the practical scenario.
Furthermore, the operating frequency is 100 GHz which corresponds to the \ac{thz} band in the spectrum, and the bandwidth is 60 MHz. In our scenario, the \ac{los} propagation distance varies between 110 and 160m. When the operating frequency is 100GHz using 256 antennas at the base station, the Rayleigh distance is approximately 100m. Thus, the signal source is always in the far-field region in our case. We can assume that the signal coming from the \ac{los} path propagates as a plane wave. In addition, depending on the environment, \ac{nlos} path can be modeled as a near-field channel and scatters propagate as a spherical wave. 

During the offline training stage, the learning rate is set as 0.001, and the mini-batch of 64 samples is used in each iteration. Also, 80\% and 20\% part of all dataset is selected as the training and validation data sets, respectively. 1000 samples that have been obtained from the receiver point at 1000 different points are used in the Monte Carlo analyses.

All the numerical results are implemented on a PC with  Intel(R) Core(TM) i9-11980HK @ 2.60GHz and Nvidia GeForce RTX 3080. Also, RIDNet is carried out by using the TensorFlow framework.

\subsection{Performance of the OMP-RIDNet}
The channel estimation performance of the OMP-RIDNet is evaluated by the \ac{nmse}
\begin{equation}
\text{NMSE} = \mathbb{E}\left(\frac{\|\mathbf{H}-\hat{\mathbf{H}}\|_{2}^{2}}{\|\mathbf{H}\|_{2}^{2}}\right).
\end{equation}
The \ac{nmse} performance of the OMP-RIDNet algorithm is compared with the \ac{omp} algorithm by varying the number of different RF chains and Q pilot matrices in \FGR{fig:nmse_rf} and \FGR{fig:nmse_pilot}, respectively. 

\begin{figure}[t!]
    \centering
    \includegraphics[width=\linewidth]{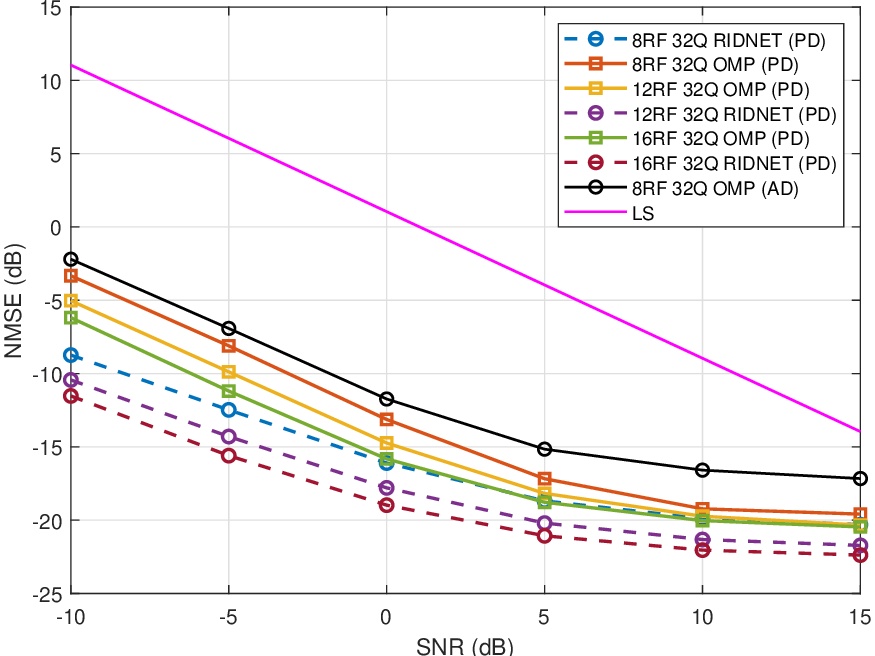}
    \caption{The NMSE performance of the proposed channel estimation scheme under $N_{RF}=8, 12, 16$}
    \label{fig:nmse_rf}
\end{figure}

\begin{figure}[ht!]
    \centering
    \includegraphics[width=\linewidth]{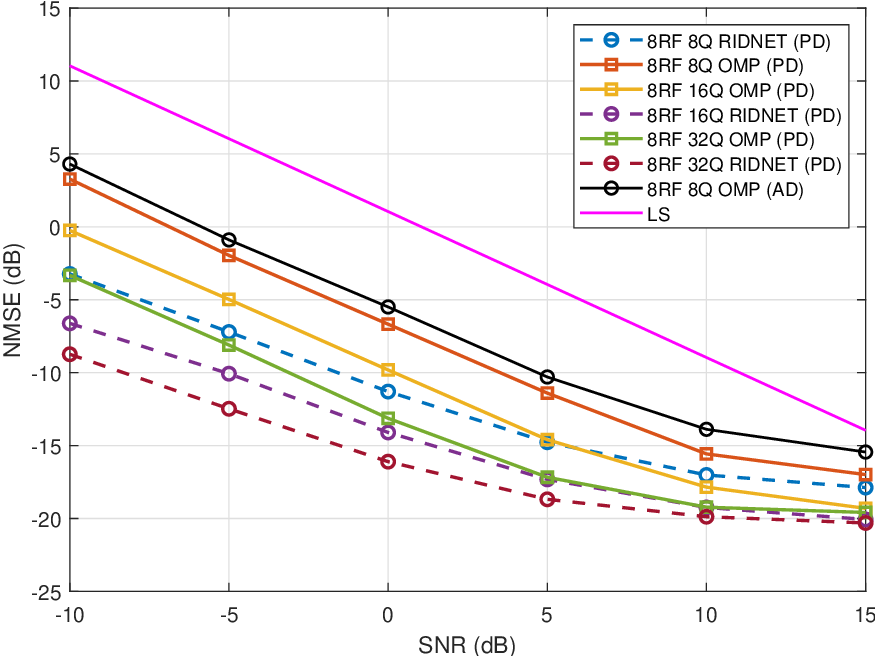}
    \caption{The NMSE performance of the proposed channel estimation scheme under $Q=8, 16, 32$}
    \label{fig:nmse_pilot}
\end{figure}

The results show that using the \ac{ad} transform matrix is a disadvantage for the channel estimation performance due to near-field components in hybrid field channels, so \ac{pd} transformation matrices with both angle and distance information should be used for the sparse representation of hybrid field channels. In \FGR{fig:nmse_rf}, the channel estimation performance of OMP-RIDNet and \ac{omp} methods is compared according to different RF chain numbers by keeping the number of pilots constant. There is a difference of 4dB at 0 SNR when 8 RF chains are used between the proposed channel estimation method and the classical \ac{omp} algorithm. In addition, the channel estimation performance obtained by using 16 RF chains can be achieved, especially at low SNRs, by using 8 RF chains with the proposed method. Thus, when using OMP-RIDNet, a significant advantage is gained from hardware complexity and power consumption.

Furthermore, the performance analysis with respect to different numbers of pilots is provided in \FGR{fig:nmse_pilot}, where the \ac{nmse} values of the \ac{omp} algorithm can be obtained using less number of pilot symbols thanks to the OMP-RIDNet channel estimation scheme. Thus, a channel estimation scheme having a lower pilot overhead can be utilized for \ac{thz} \ac{mmimo} systems.

\section{Conclusion and Future Work}
In this paper, we proposed a RIDNet-assisted two-stage channel estimation scheme for \ac{thz} \ac{mmimo} systems by taking hybrid-field channels into account. In addition, we developed a ray-tracing tool to model a THz environment with different building heights and shapes. Simulation results show that the proposed OMP-RIDNet method consistently provides better channel estimation accuracy using low pilot overhead and less number of RF chains compared to the \ac{omp} algorithm, especially for the low SNR regime. We also showed that the \ac{pd} transformation of the channel provides lower channel estimation error for hybrid-field channels, which can be represented as sparse in the \ac{pd}.

Future work will investigate one stage hybrid-field channel estimation method with lower complexity instead of a two-stage channel estimation scheme.

\section*{Acknowledgment}
This publication was made possible by the NPRP award [NPRP12S-0225-190152] from the Qatar National Research Fund, a member of The Qatar Foundation. The statements made herein are solely the responsibility of the authors. We thank to StorAIge project that has received funding from the KDT Joint Undertaking (JU) under Grant Agreement No. 101007321. The JU receives support from the European Union’s Horizon 2020 research and innovation programme in France, Belgium, Czech Republic, Germany, Italy, Sweden, Switzerland, Türkiye, and National Authority TÜBİTAK with project ID 121N350.
\balance
\bibliographystyle{IEEEtran}
\bibliography{references.bib}

\begin{thebibliography}{10}
\providecommand{\url}[1]{#1}
\csname url@samestyle\endcsname
\providecommand{\newblock}{\relax}
\providecommand{\bibinfo}[2]{#2}
\providecommand{\BIBentrySTDinterwordspacing}{\spaceskip=0pt\relax}
\providecommand{\BIBentryALTinterwordstretchfactor}{4}
\providecommand{\BIBentryALTinterwordspacing}{\spaceskip=\fontdimen2\font plus
\BIBentryALTinterwordstretchfactor\fontdimen3\font minus
  \fontdimen4\font\relax}
\providecommand{\BIBforeignlanguage}[2]{{%
\expandafter\ifx\csname l@#1\endcsname\relax
\typeout{** WARNING: IEEEtran.bst: No hyphenation pattern has been}%
\typeout{** loaded for the language `#1'. Using the pattern for}%
\typeout{** the default language instead.}%
\else
\language=\csname l@#1\endcsname
\fi
#2}}
\providecommand{\BIBdecl}{\relax}
\BIBdecl

\bibitem{sarieddeen2021overview}
H.~Sarieddeen, M.-S. Alouini, and T.~Y. Al-Naffouri, ``An overview of signal
  processing techniques for terahertz communications,'' \emph{Proceedings of
  the IEEE}, 2021.

\bibitem{akyildiz2018combating}
I.~F. Akyildiz, C.~Han, and S.~Nie, ``Combating the distance problem in the
  millimeter wave and terahertz frequency bands,'' \emph{IEEE Communications
  Magazine}, vol.~56, no.~6, pp. 102--108, 2018.

\bibitem{lu2022double}
Y.~Lu and L.~Dai, ``Double-side near-field channel estimation for extremely
  large-scale {MIMO} system,'' \emph{arXiv preprint arXiv:2205.03615}, 2022.

\bibitem{246282}
V.~Arun and H.~Balakrishnan, ``{RFocus}: Beamforming using thousands of passive
  antennas,'' in \emph{17th USENIX Symposium on Networked Systems Design and
  Implementation (NSDI 20)}, 2020, pp. 1047--1061.

\bibitem{wei2021channel}
X.~Wei and L.~Dai, ``Channel estimation for extremely large-scale massive
  {MIMO}: Far-field, near-field, or hybrid-field?'' \emph{IEEE Communications
  Letters}, vol.~26, no.~1, pp. 177--181, 2021.

\bibitem{cui2022channel}
M.~Cui and L.~Dai, ``Channel estimation for extremely large-scale {MIMO}:
  Far-field or near-field?'' \emph{IEEE Transactions on Communications},
  vol.~70, no.~4, pp. 2663--2677, 2022.

\bibitem{wei2019amp}
Y.~Wei, M.-M. Zhao, M.~Zhao, M.~Lei, and Q.~Yu, ``An {AMP-based} network with
  deep residual learning for {mmWave} beamspace channel estimation,''
  \emph{IEEE Wireless Communications Letters}, vol.~8, no.~4, pp. 1289--1292,
  2019.

\bibitem{hu2022prince}
Z.~Hu, Y.~Chen, and C.~Han, ``{PRINCE}: A pruned {AMP} integrated deep cnn
  method for efficient channel estimation of millimeter-wave and terahertz
  ultra-massive {MIMO} systems,'' \emph{arXiv preprint arXiv:2203.04635}, 2022.

\bibitem{he2018deep}
H.~He, C.-K. Wen, S.~Jin, and G.~Y. Li, ``Deep learning-based channel
  estimation for beamspace {mmWave} massive {MIMO} systems,'' \emph{IEEE
  Wireless Communications Letters}, vol.~7, no.~5, pp. 852--855, 2018.

\bibitem{ma2020sparse}
W.~Ma, C.~Qi, Z.~Zhang, and J.~Cheng, ``Sparse channel estimation and hybrid
  precoding using deep learning for millimeter wave massive {MIMO},''
  \emph{IEEE Transactions on Communications}, vol.~68, no.~5, pp. 2838--2849,
  2020.

\bibitem{chen2021hybrid}
Y.~Chen, L.~Yan, and C.~Han, ``Hybrid spherical-and planar-wave modeling and
  {DCNN}-powered estimation of terahertz ultra-massive {MIMO} channels,''
  \emph{IEEE Transactions on Communications}, vol.~69, no.~10, pp. 7063--7076,
  2021.

\bibitem{jin2019channel}
Y.~Jin, J.~Zhang, B.~Ai, and X.~Zhang, ``Channel estimation for {mmWave}
  massive {MIMO} with convolutional blind denoising network,'' \emph{IEEE
  Communications Letters}, vol.~24, no.~1, pp. 95--98, 2019.

\bibitem{chaccour2022can}
C.~Chaccour, M.~N. Soorki, W.~Saad, M.~Bennis, and P.~Popovski, ``Can terahertz
  provide high-rate reliable low latency communications for wireless {VR}?''
  \emph{IEEE Internet of Things Journal}, 2022.

\bibitem{moldovan2014and}
A.~Moldovan, M.~A. Ruder, I.~F. Akyildiz, and W.~H. Gerstacker, ``{LOS and
  NLOS} channel modeling for terahertz wireless communication with scattered
  rays,'' in \emph{IEEE Globecom Workshops (GC Wkshps)}.\hskip 1em plus 0.5em
  minus 0.4em\relax IEEE, 2014, pp. 388--392.

\bibitem{ju2019millimeter}
S.~Ju, O.~Kanhere, Y.~Xing, and T.~S. Rappaport, ``A millimeter-wave channel
  simulator {NYUSIM} with spatial consistency and human blockage,'' in
  \emph{IEEE Global Communications Conference (GLOBECOM)}.\hskip 1em plus 0.5em
  minus 0.4em\relax IEEE, 2019, pp. 1--6.

\bibitem{tarboush2021teramimo}
S.~Tarboush, H.~Sarieddeen, H.~Chen, M.~H. Loukil, H.~Jemaa, M.-S. Alouini, and
  T.~Y. Al-Naffouri, ``{TeraMIMO}: A channel simulator for wideband
  ultra-massive {MIMO} terahertz communications,'' \emph{IEEE Transactions on
  Vehicular Technology}, vol.~70, no.~12, pp. 12\,325--12\,341, 2021.

\bibitem{hur2016proposal}
S.~Hur, S.~Baek, B.~Kim, Y.~Chang, A.~F. Molisch, T.~S. Rappaport, K.~Haneda,
  and J.~Park, ``Proposal on millimeter-wave channel modeling for {5G} cellular
  system,'' \emph{IEEE Journal of Selected Topics in Signal Processing},
  vol.~10, no.~3, pp. 454--469, 2016.

\end{thebibliography}

\end{document}